\documentclass{article}

\usepackage{microtype}
\usepackage{graphicx}
\usepackage{subcaption}
\usepackage{booktabs} 

\usepackage{hyperref}

\usepackage[accepted]{icml2023}

\usepackage{amsmath}
\usepackage{amssymb}
\usepackage{mathtools}
\usepackage{amsthm}

\usepackage[capitalize,noabbrev]{cleveref}

\usepackage{array}
\newcolumntype{P}[1]{>{\centering\arraybackslash}p{#1}}

\theoremstyle{plain}

\theoremstyle{definition}

\theoremstyle{remark}

\usepackage[textsize=tiny]{todonotes}

\icmltitlerunning{AdaptiveRec: Adaptively Construct Pairs for Contrastive Learning in Sequential Recommendation}

\begin{document}

\twocolumn[
\icmltitle{AdaptiveRec: Adaptively Construct Pairs \\for Contrastive Learning in Sequential Recommendation}



\icmlsetsymbol{equal}{*}

\begin{icmlauthorlist}
\icmlauthor{Jaeheyoung Jeon}{equal,schmath}
\icmlauthor{Jung Hyun Ryu}{equal,schai}
\icmlauthor{Jewoong Cho}{equal,schmath}
\icmlauthor{Myungjoo Kang}{schmath,schai}
\end{icmlauthorlist}

\icmlaffiliation{schmath}{Department of Mathematics, Seoul National University, Seoul, Korea}
\icmlaffiliation{schai}{Interdisciplinary Program in Artificial Intelligence, Seoul National University, Seoul, Korea}

\icmlcorrespondingauthor{Myungjoo Kang}{mkang@snu.ac.kr}

\icmlkeywords{Machine Learning, ICML}

\vskip 0.3in
]



\printAffiliationsAndNotice{\icmlEqualContribution} 

\begin{abstract}
This paper presents a solution to the challenges faced by contrastive learning in sequential recommendation systems. 
In particular, it addresses the issue of false negative, which limits the effectiveness of recommendation algorithms.
By introducing an advanced approach to contrastive learning, the proposed method improves the quality of item embeddings and mitigates the problem of falsely categorizing similar instances as dissimilar. 
Experimental results demonstrate performance enhancements compared to existing systems.
The flexibility and applicability of the proposed approach across various recommendation scenarios further highlight its value in enhancing sequential recommendation systems.

\end{abstract}

\section{Introduction}

\begin{figure}[!t]
    \centerline{\includegraphics[width=\linewidth, height=5cm]{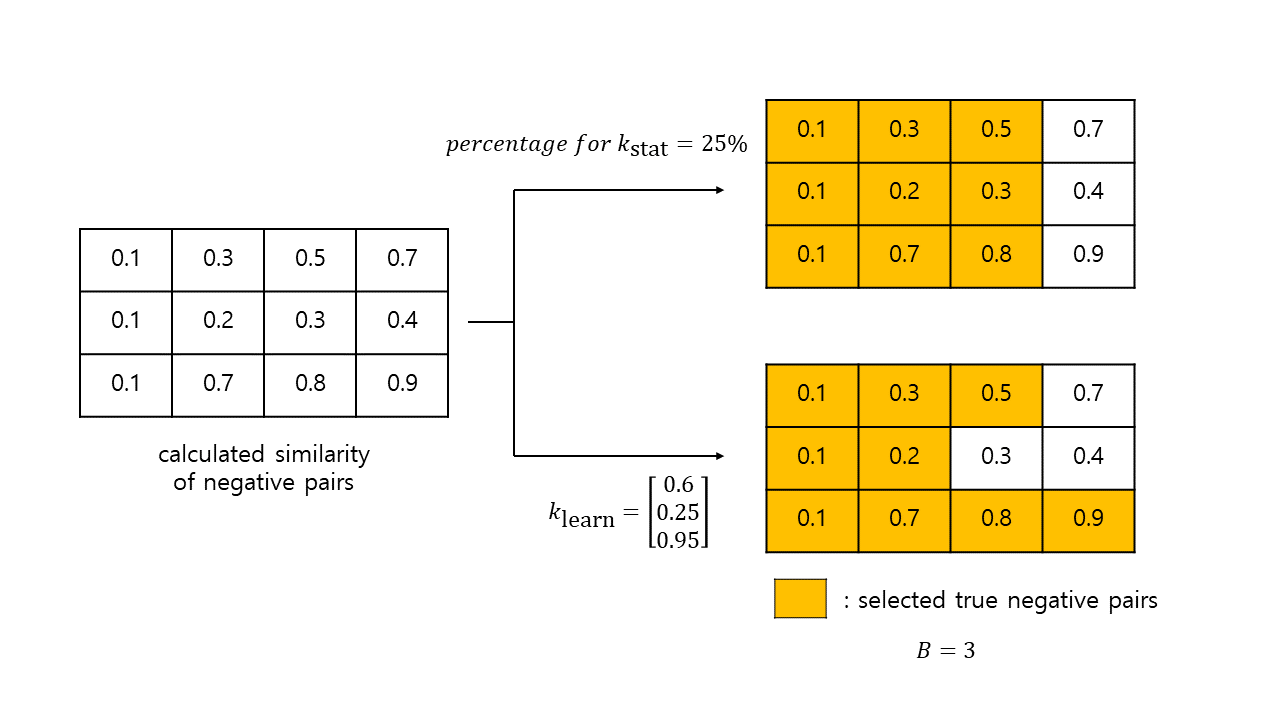}}
    \caption{The summary of our method when batch size $B$ is 3 and sorted by each row for simplicity. Total number is $B \times (2B-2)$. Each row means sequence. Our method adapts the threshold for each row to reflect the composition of the batch.The threshold can be obtained through the model's trainable parameters.}
    \label{fig:method} 
\end{figure}

In recent years, the rapid growth of online platforms and services has led to the accumulation of vast amounts of data daily.
Among the various methods of harnessing this data, recommendation systems play a prominent role. Recommendation systems are employed to identify relevant items based on user preferences and interests.
To capture the evolving user preferences over time, the field of sequential recommendation has emerged. 
Prominent examples in this domain include SASRec\cite{SASRec2018}  and BERT4Rec\cite{sun2019bert4rec}. 
In this paper, we define the problem of sequential recommendation as follows:

Let $\mathcal{U}$ be the set of users $\mathcal{U} = \{u_1, u_2, \cdots, u_{|\mathcal{U} |}\}$, and $\mathcal{V}$ be the set of items as $\mathcal{V} = \{v_1, v_2, \cdots, v_{|\mathcal{V} |}\}$.
The sequence of user-item interaction for $u_i$ is a list with chronological order, $S_i=[v_1^{u_i}, v_2^{u_i}, \cdots, v_t^{u_i}, \cdots, v_{n_{u_i}}^{u_i}]$.
Here user $u_i \in \mathcal{U}$, $v_t^{u_i} \in \mathcal{V}$, and user $u_i$ interact item $v_t^{u_i}$ in time step $t$.
The length of sequence for user $u_i$ is $n_{u_i}$, and our object is to build a model predicting the item with which user is interact in the next time step, i.e, 
\begin{equation} p(v_{n_{u_i}+1}^{u_i} = v | S_i) \end{equation}

However, the field of recommendation systems inherently faces the challenge of data sparsity.
This problem arises when the majority of elements in the user-item matrix are empty since users have not interacted with most items.
Consequently, this poses limitations on embedding the items during the modeling process.
To address this issue, previous works such as CL4SRec\cite{xie2022contrastive} and DuoRec\cite{qiu2022contrastive} have proposed methods that incorporate contrastive learning. \\
In this paper, we introduce an advanced approach to contrastive learning in sequential recommendation, called \textbf{AdaptiveRec}.
Through our experiments, we have demonstrated that the model is capable of distinguishing between negative and positive pairs.
This enable the model to adaptively identify false negatives during the training process.
Despite using a minimal additional cost, the model surpassed the performance of previous approaches.

\section{Related Work}

\subsection{Contrastive Learning}
Contrastive learning is one of self-supervised learning techniques, where the goal is to learn correct representations of data without any explicit labeling or supervision\cite{le2020contrastive}.
In contrastive learning, the algorithm is trained to compare and contrast different samples of data to learn the underlying patterns and relationships between them.
The basic idea is to take two different data points and generate two different representations for each data point\cite{oord2018representation}.
The algorithm aims to learn to differentiate between representations of the same data point, referred to as a positive pair, and representations of different data points, known as a negative pair, using a similarity metric.
The training process involves optimizing the model to maximize the similarity between the representations of positive pairs while minimizing the similarity between the representations of negative pairs.
This way, the model learns to capture the meaningful differences and similarities between the data points, which can then be used for downstream tasks such as classification, clustering, or retrieval.

Contrastive learning can be mathematically formulated using the InfoNCE (Normalized Mutual Information Neural Estimation)\cite{wu2022infocse} loss function, which is used to maximize the similarity between positive pairs and minimize the similarity between negative pairs.
The InfoNCE loss is defined as:
\begin{equation}\label{eq:infoNCE}
\begin{alignedat}{2}
   \ell_\mathsf{NCE} &=\sum \limits _{i \in I} \Bigg[-\log \cfrac{e^{\text{sim}(f(s_i), f(s_i^+))}}{\sum \limits _{j \in M(i)} e^{\text{sim}(f(s_i), f(s_j))}} \Bigg], \\
    M(i) &= \{i^+, j \mid j \in N(i) \}, \\
    N(i) &= negative\; pairs\; of\; i .
\end{alignedat}
\end{equation}

\subsection{Contrastive Learning in Sentence Embedding}
SimCSE\cite{gao2021simcse} is a framework for contrastive sentence embedding that improves the quality of sentence representations using pretrained embeddings. 
By applying contrastive learning principles, it achieves enhanced alignment and uniformity within the embeddings by bringing similar samples closer and separating dissimilar ones.
DiffCSE \cite{chuang2022diffcse} is an extention of SimCSE \cite{gao2021simcse} that aims to overcome limitations.
It addresses these limitations by incorporating additional augmentation methods and promoting equvariant contrastive learning for sentence embeddings \cite{dangovski2021equivariant}.
PromCSE\cite{jiang2022improved} is a framework that utilizes the concept of hard negatives, which are negative pairs that are difficult to differentiate from positive pairs. By leveraging the presence of these challenging negative pairs, PromCSE aims to improve discrimination in supervised learning.

\subsection{Contrastive Learning in Sequential Recommendation}
Contrastive Learning for Sequential Recommendation\cite{xie2022contrastive} is proposed as a solution to address the significant issue of data sparsity in recommendation systems by incorporating contrastive learning techniques.
CL4SRec\cite{xie2022contrastive} focuses on constructing pairs with different viewpoints, where positive pairs carry semantic information.
The paper introduces three data augmentation techniques at the sequence level to obtain more effective user representations. These augmentation methods include item cropping, which involves cropping a certain interval within the sequence, item masking, which masks several items within the sequence, and item reordering, which rearranges the order of items within a specific interval in the sequence.
By applying these data augmentation techniques, CL4SRec\cite{xie2022contrastive} performs contrastive learning by creating positive pairs from the original sequence.
These augmentation methods closely resemble the data augmentation approaches commonly used in the field of computer vision.\cite{chen2020simple}
This integration of contrastive learning and sequence-based data augmentation in CL4SRec\cite{xie2022contrastive} contributes to advancements in addressing data sparsity and enhancing sequential recommendation systems.

DuoRec\cite{qiu2022contrastive} model combines two forms of contrastive loss.
Firstly, it utilizes unsupervised augmentation through dropout-based model-level augmentation to create positive pairs.
It emphasizes the importance of pairing samples under the assumption that their meaning remains unchanged.
The paper also highlights that semantic preservation alone in CL4SRec\cite{xie2022contrastive} may not be sufficient.
In this regard, the proposed DuoRec\cite{qiu2022contrastive} model heavily incorporates the feature-level augmentation through dropout, as suggested by SimCSE\cite{gao2021simcse}. Additionally, DuoRec\cite{qiu2022contrastive} employs supervised positive sampling, where pairs are created by considering sequences with the same target item as positive samples.
This approach enhances the model's ability to capture the relationships between sequences with shared target items.
By combining both unsupervised and supervised contrastive learning, DuoRec\cite{qiu2022contrastive} aims to improve the effectiveness of recommendation models in capturing semantic information and considering item-level relationships.

\subsection{Alignment and Uniformity}
\label{sec:uni_ali}
In order to point out how well features are represented, previous work has suggested two key properties; alignment and uniformity\cite{wang2020understanding}.
Alignment measures how well paired positive instances are closely represented, typically computed as the average distance between them.
On the other hand, uniformity assesses how well the feature distribution is uniformly spread, often quantified using a Gaussian kernel on a hypersphere.
Given the data distribution $p_{data}$ and a positive pair distribution $p_{pos}$, alignment and uniformity is defined as : 

\begin{equation}\label{eq:align} \ell_{\text{align}} \triangleq \mathbb{E}_{(x,x^+) \sim p_{\text{pos}}} \| f(x) - f(x^+) \|^2 \end{equation}
\begin{equation}\label{eq:uniform} \ell_{\text{uniform}} \triangleq 
\log \mathbb{E}_{(x,y) \sim p_{\text{data}}} e^{-2 \| f(x) - f(y) \|^2} \end{equation}

Minimizing both metrics indicates an improved performance.
Minimizing $\ell_{\text{align}}$ in equation \ref{eq:align} encourages the learned representations of samples $x$ and $x^+$, sampled from $p_{\text{pos}}$, to be closer together.
Minimizing $\ell_{\text{uniform}}$ in equation \ref{eq:uniform} encourages the samples from $p_{\text{data}}$ to be uniformly represented.
By considering these two objectives, the goal is to enhance the proximity of positive pairs while promoting uniform representation across the data distribution.

\section{Method}

\subsection{AdaptiveRec}
We introduce Adaptive Sequential Recommendation, named \textbf{AdaptiveRec}, which employs an adaptive thresholding method to distinguish between negative and positive pairs.

Previous works \cite{xie2022contrastive, qiu2022contrastive} addressed item embedding in sequential recommendation through contrastive learning primarily focused on enhancing positive samples.
However, defining dissimilar sequences posed challenges as the recommendation domain exhibits high sparsity \cite{yao2020self, zhou2020s3} and significant variation in item popularity. 
It is difficult to establish meaningful distinctions between sequences. 
To tackle this issue, we approach the problem from the perspective of analyzing methods for identify true negative samples, which is the main idea of AdaptiveRec.
Specifically, we define the relationship between item sequences based on similarity and incorporated this relationship into the contrastive loss.

The contrastive learning and batch-based positive and negative pair selection methods employed in SimCLR \cite{chen2020simple} were both simple and intuitive.
With batch size $B$, the number of sequences is augmented to $2B$.
Each sequence considers the remaining $2B-2$ sequences, excluding itself and its own augmented versions, as negative pairs.
However, these approaches do have inherent limitations, such as the occurrence of false negatives, where negative pairs can be semantically similar.
Therefore, constructing well-designed negative pairs holds great importance in contrastive learning.

\begin{figure}[!t]
    \centering
    \begin{subfigure}{0.495\columnwidth}
        \centering
        \includegraphics[width=\linewidth]{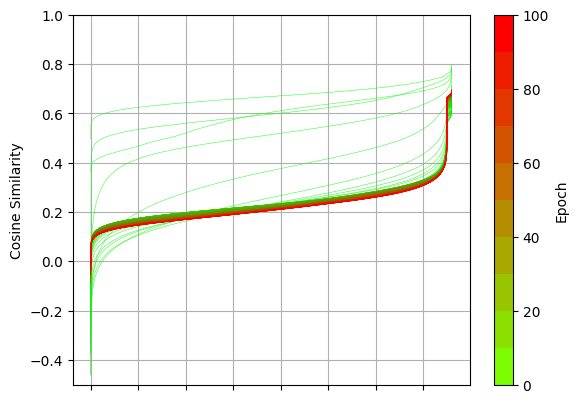}
        \caption{CL4SRec}
        \label{cl4srec}
    \end{subfigure}
    \hfill
    \begin{subfigure}{0.495\columnwidth}
        \centering
        \includegraphics[width=\linewidth]{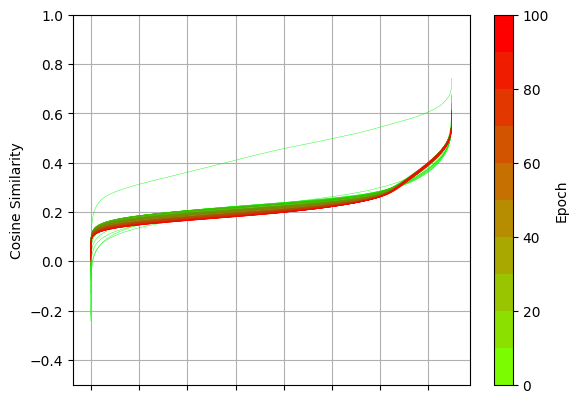}
        \caption{DuoRec}
        \label{duorec}
    \end{subfigure}
\caption{Calculated Similarity of CL4SRec and DuoRec during Training. As training preceed, graph turns from green to red. Similaritiy distribution is sorted along with every epochs. As graphs turn to red, the distribution become sharp after threshold.  }
\label{fig:similarity}
\end{figure}

As illustrated in Fig \ref{fig:similarity}, we assume that the model learns the ability to distinguish \textit{false negative} pairs from genuine negative pairs. Despite only showing the values of negative pairs in the graph, we observe that the similarity value increases sharply beyond a certain threshold.
More detailed discussions on the matter can be found in Sec. \ref{App_sec:feature_distribution_and_similarity}.

We construct negative samples based on the similarity with other samples within the batch. 
In a scenario where the batch size is $B$, we first form the positive pairs as suggested in DuoRec \cite{qiu2022contrastive}. 
The negative samples are then selected from the remaining $2B-2$ samples, considering only those with similarity below a certain threshold.

We refer to this threshold as \textbf{\textit{SimThres}}, and it allows dynamic negative pair thresholding based on the chosen \textit{SimThres} value. 
We propose adaptive learnable thresholding method, \textit{SimThres}, in Section \ref{sc_learnable_simthres}.
These approaches enable us to define true negative pairs by considering the similarity threshold, allowing better exploration of the dissimilarity between item sequences in the context of contrastive learning.

\subsection{Determining the \textbf{\textit{SimThres}}}
For user $u_i \in \mathcal{U}$ and item $v_j \in \mathcal{V}$, the sequence of user-item interaction for $u_i$ is denoted as $s_i=[v_1^{u_i}, v_2^{u_i}, \cdots, v_t^{u_i}, \cdots, v_{n_{u_i}}^{u_i}]$.
Let $\mathcal{S}=\{s_i \; | \; i=1,\cdots,|\mathcal{U}|\}$ represent the set of user-item interaction.
Also the neural network model is $f$ and $s_i^+$ is the augmented positive pair of $s_i$.

For each $s_i$, we consider an ordered set, $\{s_{i,j}\}$, where $j$ is the order index, sorted based on the similarity value with respect to $s_i$.
For instance, 
\begin{equation}
    s_{i,1} = \underset{(s_j \in \mathcal{S}) \wedge (s_j \neq s_i,s_i^+)}{\arg\min} \text{sim} (f(s_i),f(s_j)).\end{equation}

We aim to identify true negative pairs for each $u_i$ using values from this set that do not exceed a certain threshold.
In other words, the negative samples are given by 
\begin{equation} \label{eq:negative_sample}
    \{s_{i,j} \mid \text{sim} (f(s_i; \theta),f(s_{i,j}; \theta)) < \textbf{k} \}.
\end{equation}
We refer to the threshold denoted as $\mathbf{k}$ in Equation \ref{eq:negative_sample} as \textit{SimThres}.

\subsubsection{Statistical \textit{SimThres}}
In a statistical manner, we set the \textit{SimThres} based on the similarity values with the entire set of users.
The $k_{stat}$ is representing the percentile of negative pairs. 
We defined the negative pairs for each $i$ in the batch as the pairs, $(s_i,s_j)$, that fall within the lower $k_{stat}$ percentile.
We only use that pairs as negative pairs in contrastive learning loss.
Through experimentation, we found that setting the $k_{stat}$ to the 90\%.

\subsubsection{Learnable \textit{SimThres} where statistically regularized} \label{sc_learnable_simthres}

\begin{figure}
     \centering
     \begin{subfigure}[]{0.9\columnwidth}
         \centering
         \includegraphics[width=0.9\textwidth]{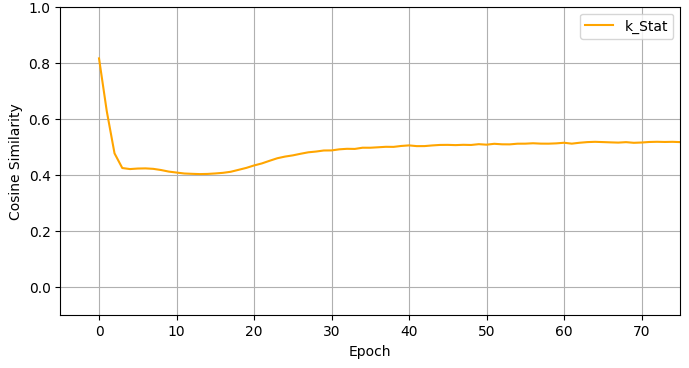}
         \caption{Statistical \textit{SimThres}}
     \end{subfigure}
     \hfill
     \begin{subfigure}[b]{0.9\columnwidth}
         \centering
         \includegraphics[width=0.9\textwidth]{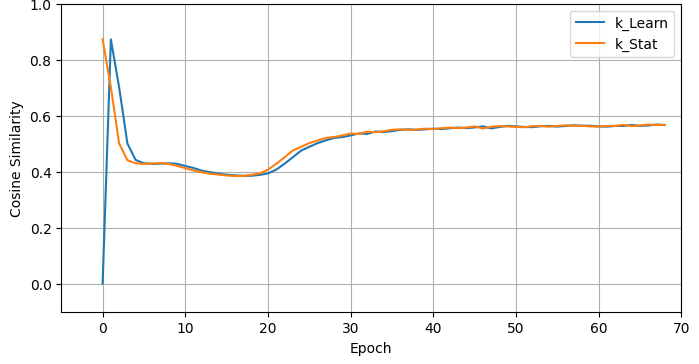}
         \caption{Learnable \textit{SimThres} where statistically regularized}
     \end{subfigure}
     \caption{The \textit{SimThres}, which changes as the training progresses, is converging to a stable value. It shows similar trends and convergence patterns in both cases, but the final converged values differ.}
     \label{fig:simthres_k}
\end{figure}

We observe that $k_{\text{stat}}$ converges well as training progressed in Fig \ref{fig:simthres_k}. However we also observe that the lower 90\% similarity score is not appropriate for the practical batch-wise training.
The global 90\% may not align with the local 90\% in batch-wise training.
Furthermore, even for the same sequence, the composition of the batches it belongs to can vary across epochs. As a result, the similarity calculations can also differ accordingly.
Therefore, even for the same sequence, customized thresholds,$k_\text{learn}$ based on batch composition are necessary.

In order to obtain the value of $k_{\text{learn}}$ as a trainable parameter, we introduce an additional submodel $g$ that enables us to set the \textit{SimThres} adaptively as follow:
\begin{equation} k_{\text{learn}}=g(\{f(s_i)\}_{i=1}^B). \end{equation}
, where $g : B \times \mathbb{R}^{D} \rightarrow B$, $B$ is batch size and $D$ is the dimension of feature.

We employ a straightforward Multi-Layer Perceptron (MLP) architecture for submodel $g$. 
By incorporating this trainable submodel, we empower our methodology to learn and adapt the value of $k_{\text{learn}}$ for each sequence in every batch during the training process, enhancing the flexibility and effectiveness of our approach in capturing the desired negative sample characteristics.

To control the value of \textit{SimThres} and avoid the collapsing phenomenon of $k_{\text{learn}}$ (Appendix \ref{App_sec:collapse_phenomenon}), we introduce a regularization term to our methodology.
The regularization term, denoted as $\mathcal{L}_{reg}$, is implemented by adding a loss term of the form $\mathcal{L}_{reg}=\|g(\{f(s_i)\}_{i=1}^B)-\overline{k_{stat}}\|_2^2$, where $\overline{k_{stat}}$ is the lower $k_{stat}(\%)$ value of similarity in before epoch and serves as the control value.
This additional term allows us to enforce a desired level of regularization on the value of $g(\{f(s_i)\}_{i=1}^B)$.
We refer to the resulting value $g(\{f(s_i)\}_{i=1}^B)$ as $k_{\text{learn}}$, representing the regulated and learned threshold.
By incorporating this regularization mechanism, we aim to achieve a balance between adapting the threshold value through training and maintaining consistency with the desired statistical properties represented by $k_{\text{stat}}$.
$M^{-}(i)$ is the set of $j$ such that ${\text{sim}(f(s_i), f(s_j))}$ is lower than customized threshold for $i$.
$N(i)$ is the set of indices for negative pairs in-batch negatives as previous works defined.
So, our loss function is as follows:
\begin{equation}\label{eq:infoNCE}
\begin{alignedat}{3}
    \mathcal{L}_{learn} &=\sum \limits _{i=1}^{2B} \Bigg[-\log \cfrac{e^{\text{sim}(f(s_i), f(s_i^+))}}{\sum \limits _{j \in M^{-}(i)\cup\{i^+\}} e^{\text{sim}(f(s_i), f(s_j))}} \Bigg], \\
    M^{-}(i) &= \,\{\, j \in N(i) \mid {\text{sim}(f(s_i), f(s_j))}\leq k_\text{learn}[i] \,\}, \\
    N(i) &= \{\, 1,2, ..., 2B-2 \, \}.
\end{alignedat}
\end{equation}

\subsubsection{Positive Sampling Strategy}
We assume that if the model successfully identifies pairs with a similarity lower than \textit{SimThres} as negative pairs, it implies that pairs with higher values can be considered positive pairs. Therefore, we proceed to label pairs with similarity values greater than \textit{SimThres} as positive pairs and conduct further training, where $s_{i^+}$ indicates the augmented sequence from $s_i$.
To assign weights to the labels, we assign a higher weight to label $s_{i^+}$ and a slightly smaller weight to other positive pairs. The sum of the weights is equal to $1$.
So, our loss function considering positive sampling is as follows:
\begin{equation}\label{eq:infoNCE}
    \begin{alignedat}{6}
    \mathcal{L}_{learn} =\sum \limits _{i=1}^{2B} &\Bigg[-\log \frac{\frac{1}{2}e^{\text{sim}(f(s_i), f(s_i^+))}}{\sum \limits _{j \in M^{-}(i)\cup\{i^+\}} e^{\text{sim}(f(s_i), f(s_j))}} \, + \\ 
   &\sum \limits _{j^+ \in M^{+}(i)}-\log \frac{\frac{1}{2 \mid M^{+}(i) \mid}e^{\text{sim}(f(s_i), f(s_{j^+}))}}{\sum \limits _{j \in M^{-}(i)\cup\{i^+\}} e^{\text{sim}(f(s_i), f(s_j))}}\Bigg], \\
    M^{-}(i) &= \,\{\, j \in N(i) \mid {\text{sim}(f(s_i), f(s_j))}\leq k_\text{learn}[i] \,\}, \\
    M^{+}(i) &= \,\{\, j \in N(i) \mid {\text{sim}(f(s_i), f(s_j))} > k_\text{learn}[i] \,\}, \\
    \mathcal{L}_{cl}\; &= \mathcal{L}_{learn}\; + \lambda\mathcal{L}_{reg}. \\
    \mathcal{L}_{total}\; &= \mathcal{L}_{basic}\; + \lambda_{cl}\mathcal{L}_{cl}. \\
    \end{alignedat}
\end{equation}

In all our experiments, AdaptiveRec trained with the loss function in Eq.\, \ref{eq:infoNCE}.

\section{Experiment}

\begin{table*}[h]
\caption{The overall performance of AdaptiveRec is evaluated using three metrics: NDCG, MRR, and Recall. The best scores are indicated in bold. Improvement percentiles are calculated based on DuoRec. AdaptiveRec demonstrates improvements across all the overall metrics..}
\vskip 0.15in
\begin{center}
\begin{small}
\begin{sc}
\resizebox{\textwidth}{!}{\begin{tabular}{c|cccccc|cccccc}
\toprule
Dataset                & \multicolumn{6}{c|}{ML-1M}                                                                                & \multicolumn{6}{c}{Amazon Beauty}                                                  \\
\midrule
Metric & \multicolumn{2}{c}{NDCG@}         & \multicolumn{2}{c}{MRR@}          & \multicolumn{2}{c|}{Recall@}      & \multicolumn{2}{c}{NDCG@} & \multicolumn{2}{c}{MRR@} & \multicolumn{2}{c}{Recall@} \\
                       & 5               & 10              & 5               & 10              & 5               & 10              & 5           & 10          & 5           & 10         & 5            & 10           \\
\midrule
CL4SRec                & 0.0726          & 0.0955          & 0.0571          & 0.0665          & 0.12            & 0.1916          &             &             &             &            &              &              \\
DuoRec                 & 0.1074          & 0.1348          & 0.089           & 0.0996          & 0.1632          & 0.2485          & 0.0329      & 0.0421      & 0.0252      & 0.0289     & 0.0564       & 0.0852       \\
\midrule
\begin{tabular}[c]{@{}c@{}}AdaptiveRec\\ (ours)\end{tabular}     & \textbf{0.1135} & \textbf{0.1411} & \textbf{0.0945} & \textbf{0.1058} & \textbf{0.1738} & \textbf{0.2611} & \textbf{0.0335}      & \textbf{0.0427}      & \textbf{0.0257}      & \textbf{0.0295}     & \textbf{0.0574}       & \textbf{0.0857}       \\
Improv.                & 5.68\%          & 4.67\%          & 6.12\%          & 6.22\%          & 6.5\%           & 5.07\%          & 1.82\%      & 1.43\%      & 1.98\%      & 2.08\%     & 1.78\%       & 0.59\% \\
\bottomrule
\end{tabular}}
\end{sc}
\end{small}
\end{center}
\vskip -0.1in
\label{table:dataset}
\end{table*}

In our experiments, we demonstrate improvements in overall performance across different datasets and augmentation settings. Furthermore, we assess the embedding quality of our model based on uniformity and alignment. Our results confirm that our model surpasses existing models in terms of both performance and embedding quality.

\subsection{Setup}
\subsubsection{Dataset}
The dataset used in this study is MovieLens-1M \cite{harper2015movielens} and Amazon Beauty \cite{mcauley2015image}.
ML-1M consists of ratings of movies and Amazon-Beauty consists for amazon beauty products and ratings.
Both datasets are frequently used for Sequential Recommendation System.
Both datasets provide valuable resources for training and evaluating recommendation models.
The dataset comprises sequential data for individual users, capturing the items they purchased in a chronological order. In our experimental setup, we designate the last item in the sequence as the test data, the item immediately preceding it as the validation data, and the preceding items as the training data. To train the model, we adopt a masked language modeling approach inspired by BERT \cite{devlin2018bert}. This approach involves masking specific items in the sequential list and training the model to predict the masked items.

\subsubsection{Metric}
The evaluation method utilized in this study is the normalized discounted cumulative gain (NDCG), a ranking-based metric \cite{he2017neural}. It entails ranking the top items predicted by the model based on their perceived preference and comparing this ranking to the actual ranking of preferred items. A higher NDCG value, closer to 1, indicates superior performance.
Additionally, we employ Mean Reciprocal Rank (MRR) and Recall metrics to rigorously evaluate the performance of the trained model.
All metrics mean superior performance, as close to 1.
Our metric values may vary depending on the item selection criteria, such as using the complete item pool, a random subset of 100 items, or the 100 popular items.

\subsubsection{Baselines}
We set our control group models as CL4SRec \cite{xie2022contrastive} and DuoRec \cite{qiu2022contrastive}. Our method can be applied in both models. We set DuoRec as baseline and apply our methods. DuoRec has 3 augmentation strategies for contrastive learning: supervised, unsupervised and mixed one, which are denoted as SU, UN, US\_X in Table \ref{table:aug_setting}.

\subsection{Performance}
In this experiment, we divided the evaluation setting into two categories to represent the model's performance with high reliability. First in Table \ref{table:dataset}, we conduct experiment in ML-1M \cite{harper2015movielens} and Amazon Beauty datasets \cite{mcauley2015image}. All metrics are performed in whole setting. At every metric our method outperform previous models. Second, in Table \ref{table:aug_setting}, we compare our method to DuoRec with every augmentation strategy and every candidate setting. Our methods outperform in three candidate setting.

\begin{figure}[h]
    \centerline{\includegraphics[width=\linewidth, height=5cm]{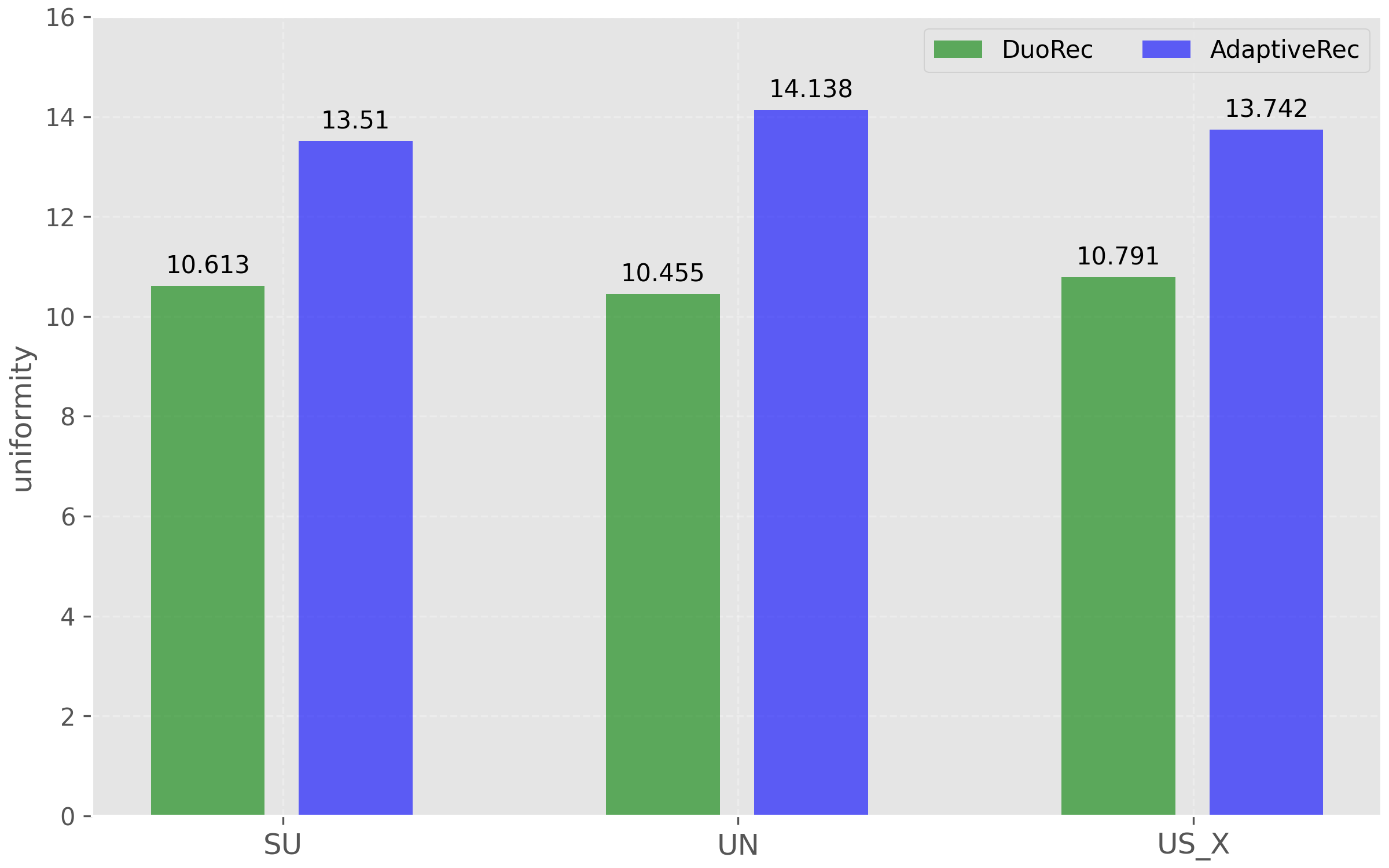}}
    \caption{We measured the uniformity of DuoRec and AdaptiveRec. In the context of uniformity, we inverted the sign to enhance visual intuitiveness. Originally, lower negative values indicated better uniformity. However, in our representation, higher positive values in the visualization indicate better uniformity. Across three augmentation settings, AdaptiveRec outperformed DuoRec in terms of uniformity. }
    \label{fig:uniformity} 
\end{figure}

\begin{figure}[]
    \centerline{\includegraphics[width=0.8\linewidth]
    {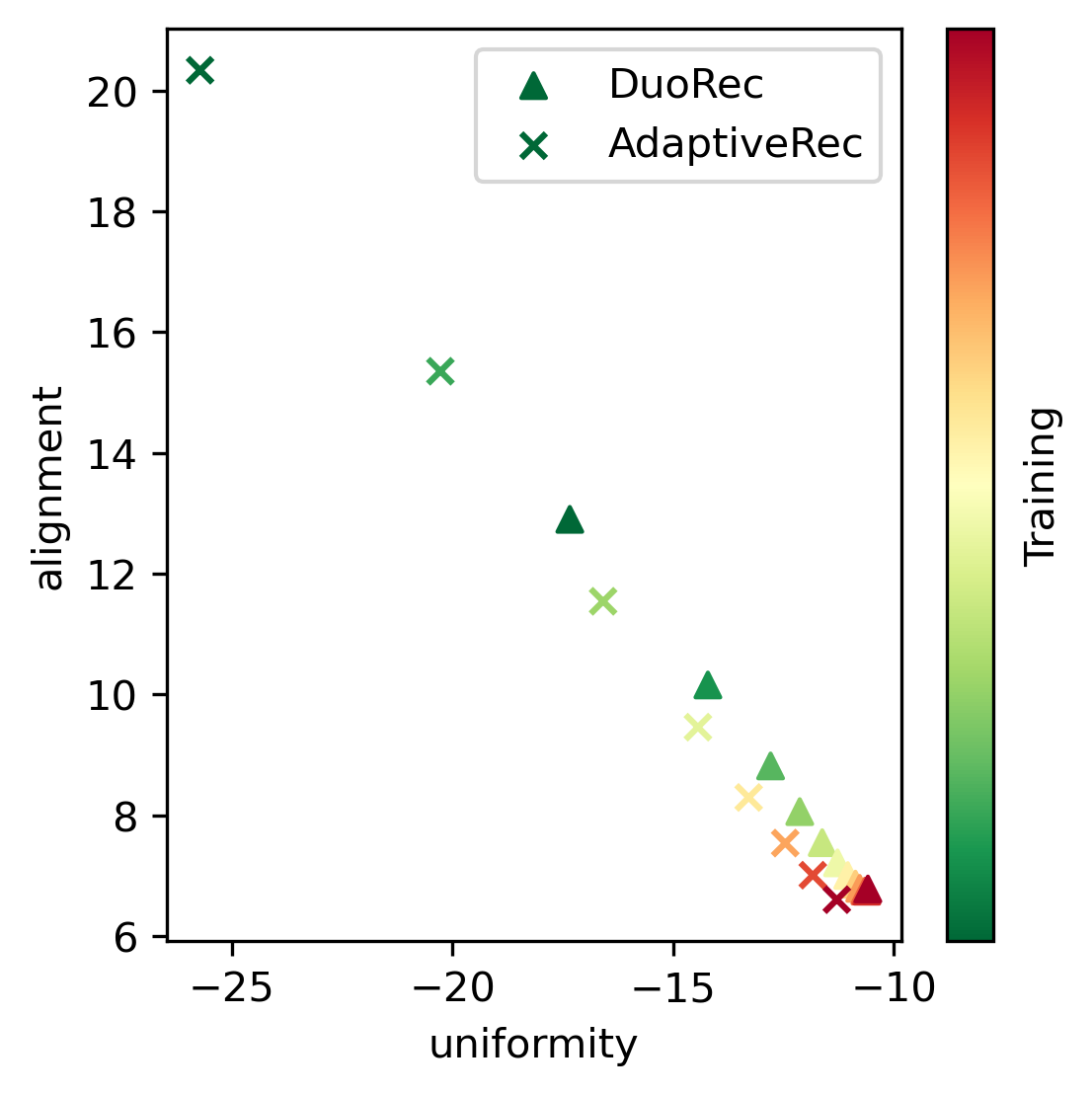}}
    \caption{The comparison of observed changes in alignment and uniformity based on the direction of training progression, green to red. When compared to DuoRec, AdaptiveRec generally exhibits lower values in terms of both alignment and uniformity.}
    \label{fig:unif_align} 
\end{figure}

\begin{table*}[t]
\caption{We do experiments with 3 candidates setting, whole, popular and random. AdaptiveRec got best score in every candidates setting. Metric is NDCG@10. }
\vskip 0.15in
\begin{center}
\begin{small}
\begin{sc}
\begin{tabular}{c|ccc|ccc|ccc}
\toprule
                   & \multicolumn{3}{c|}{whole}                    & \multicolumn{3}{c|}{popular}                  & \multicolumn{3}{c}{random}              \\
                   & su           & un           & us\_x           & su                 & un     & us\_x           & su                & un              & us\_x        \\ 
\midrule
Duorec             & 0.1386       & 0.1357       & 0.1389          & 0.0405             & 0.0442 & 0.0444          & 0.5621            & 0.5641          & 0.5668 \\
AdaptiveRec (ours) & 0.1367       & 0.1396       & \textbf{0.1410} &    0.0444          & 0.0433 & \textbf{0.0463} &  \textbf{0.5680}  & 0.5675          & 0.5643       \\ 
\bottomrule
\label{table:aug_setting}
\end{tabular}
\end{sc}
\end{small}
\end{center}
\vskip -0.1in
\end{table*}

\subsection{Uniformity and Alignment}
As a metric for evaluating the model's performance beyond traditional metrics, we quantified the embedding quality using measures of uniformity and alignment, depicted in Section.\ref{sec:uni_ali}.
First, in Fig \ref{fig:uniformity}, we measured the uniformity of our model and the DuoRec. We measured for all three data augmentation methods, and our model consistently showed higher performances in all three settings compared to the baseline. Bigger values indicate better performance, Fig \ref{fig:uniformity}.
Secondly, in Fig \ref{fig:unif_align}, to visualize both uniformity and alignment together, we plotted the uniformity values on the x-axis and alignment values on the y-axis. We set the color of each point to transition from green to red as the models were trained. This allowed us to observe the change in both uniformity and alignment as the models progressed. Lower values indicate better performance. Uniformity and Alignment have shown trade-off relationship in previous studies. Our model shows better trade-off in Fig \ref{fig:unif_align}.

\subsection{Analysis of Feature Distribution and Similarity}
\label{App_sec:feature_distribution_and_similarity}
While the minimizing contrastive learning loss, the positively augmented pair get closer and negative pairs become further apart from each other.
As shown in Fig \ref{fig:similarity}, the features of item sequences become further apart from each other and are predominantly distributed at low cosine similarity values.
Although Fig \ref{fig:similarity} only displays the cosine similarity of negative pairs, as the training progresses, the distribution of high cosine similarity values becomes more evident.
We believe that these high cosine similarity values correspond to semantically similar pairs but are falsely classified as negative pairs, resulting in \textit{false negatives}.

\section{Conclusion}

This paper proposes two important contributions to address the issues of sequential recommendation systems. Firstly, it introduces an advanced approach to contrastive learning, improving item embedding quality and mitigating the false negative problem caused by contrastive learning. The experimental results show consistent performance improvements compared to existing systems. The proposed approach, adaptive and dynamic thresholding method, is also flexible and applicable to various recommendation scenarios. Overall, this paper provides a valuable solution to enhance sequential recommendation systems.

\section*{Acknowledgements}
This work was support by the NRF grant [2012R1A2C3010887] and the MSIT/IITP [1711117093, 2021-0-00077, 2021-0-01343, Artificial Intelligence Graduate School Program(SNU)].

\nocite{langley00}

\bibliography{reference}

\begin{thebibliography}{19}
\providecommand{\natexlab}[1]{#1}
\providecommand{\url}[1]{\texttt{#1}}
\expandafter\ifx\csname urlstyle\endcsname\relax
  \providecommand{\doi}[1]{doi: #1}\else
  \providecommand{\doi}{doi: \begingroup \urlstyle{rm}\Url}\fi

\bibitem[Chen et~al.(2020)Chen, Kornblith, Norouzi, and Hinton]{chen2020simple}
Chen, T., Kornblith, S., Norouzi, M., and Hinton, G.
\newblock A simple framework for contrastive learning of visual
  representations.
\newblock In \emph{International conference on machine learning}, pp.\
  1597--1607. PMLR, 2020.

\bibitem[Chuang et~al.(2022)Chuang, Dangovski, Luo, Zhang, Chang,
  Solja{\v{c}}i{\'c}, Li, Yih, Kim, and Glass]{chuang2022diffcse}
Chuang, Y.-S., Dangovski, R., Luo, H., Zhang, Y., Chang, S.,
  Solja{\v{c}}i{\'c}, M., Li, S.-W., Yih, W.-t., Kim, Y., and Glass, J.
\newblock Diffcse: Difference-based contrastive learning for sentence
  embeddings.
\newblock \emph{arXiv preprint arXiv:2204.10298}, 2022.

\bibitem[Dangovski et~al.(2021)Dangovski, Jing, Loh, Han, Srivastava, Cheung,
  Agrawal, and Solja{\v{c}}i{\'c}]{dangovski2021equivariant}
Dangovski, R., Jing, L., Loh, C., Han, S., Srivastava, A., Cheung, B., Agrawal,
  P., and Solja{\v{c}}i{\'c}, M.
\newblock Equivariant contrastive learning.
\newblock \emph{arXiv preprint arXiv:2111.00899}, 2021.

\bibitem[Devlin et~al.(2018)Devlin, Chang, Lee, and Toutanova]{devlin2018bert}
Devlin, J., Chang, M.-W., Lee, K., and Toutanova, K.
\newblock Bert: Pre-training of deep bidirectional transformers for language
  understanding.
\newblock \emph{arXiv preprint arXiv:1810.04805}, 2018.

\bibitem[Gao et~al.(2021)Gao, Yao, and Chen]{gao2021simcse}
Gao, T., Yao, X., and Chen, D.
\newblock Simcse: Simple contrastive learning of sentence embeddings.
\newblock \emph{arXiv preprint arXiv:2104.08821}, 2021.

\bibitem[Harper \& Konstan(2015)Harper and Konstan]{harper2015movielens}
Harper, F.~M. and Konstan, J.~A.
\newblock The movielens datasets: History and context.
\newblock \emph{Acm transactions on interactive intelligent systems (tiis)},
  5\penalty0 (4):\penalty0 1--19, 2015.

\bibitem[He et~al.(2017)He, Liao, Zhang, Nie, Hu, and Chua]{he2017neural}
He, X., Liao, L., Zhang, H., Nie, L., Hu, X., and Chua, T.-S.
\newblock Neural collaborative filtering.
\newblock In \emph{Proceedings of the 26th international conference on world
  wide web}, pp.\  173--182, 2017.

\bibitem[Jiang et~al.(2022)Jiang, Zhang, and Wang]{jiang2022improved}
Jiang, Y., Zhang, L., and Wang, W.
\newblock Improved universal sentence embeddings with prompt-based contrastive
  learning and energy-based learning.
\newblock In \emph{Findings of the Association for Computational Linguistics:
  EMNLP 2022}, pp.\  3021--3035, 2022.

\bibitem[Kang \& McAuley(2018)Kang and McAuley]{SASRec2018}
Kang, W.-C. and McAuley, J.
\newblock Self-attentive sequential recommendation.
\newblock In \emph{2018 IEEE international conference on data mining (ICDM)},
  pp.\  197--206. IEEE, 2018.

\bibitem[Le-Khac et~al.(2020)Le-Khac, Healy, and Smeaton]{le2020contrastive}
Le-Khac, P.~H., Healy, G., and Smeaton, A.~F.
\newblock Contrastive representation learning: A framework and review.
\newblock \emph{Ieee Access}, 8:\penalty0 193907--193934, 2020.

\bibitem[McAuley et~al.(2015)McAuley, Targett, Shi, and Van
  Den~Hengel]{mcauley2015image}
McAuley, J., Targett, C., Shi, Q., and Van Den~Hengel, A.
\newblock Image-based recommendations on styles and substitutes.
\newblock In \emph{Proceedings of the 38th international ACM SIGIR conference
  on research and development in information retrieval}, pp.\  43--52, 2015.

\bibitem[Oord et~al.(2018)Oord, Li, and Vinyals]{oord2018representation}
Oord, A. v.~d., Li, Y., and Vinyals, O.
\newblock Representation learning with contrastive predictive coding.
\newblock \emph{arXiv preprint arXiv:1807.03748}, 2018.

\bibitem[Qiu et~al.(2022)Qiu, Huang, Yin, and Wang]{qiu2022contrastive}
Qiu, R., Huang, Z., Yin, H., and Wang, Z.
\newblock Contrastive learning for representation degeneration problem in
  sequential recommendation.
\newblock In \emph{Proceedings of the fifteenth ACM international conference on
  web search and data mining}, pp.\  813--823, 2022.

\bibitem[Sun et~al.(2019)Sun, Liu, Wu, Pei, Lin, Ou, and
  Jiang]{sun2019bert4rec}
Sun, F., Liu, J., Wu, J., Pei, C., Lin, X., Ou, W., and Jiang, P.
\newblock Bert4rec: Sequential recommendation with bidirectional encoder
  representations from transformer.
\newblock In \emph{Proceedings of the 28th ACM international conference on
  information and knowledge management}, pp.\  1441--1450, 2019.

\bibitem[Wang \& Isola(2020)Wang and Isola]{wang2020understanding}
Wang, T. and Isola, P.
\newblock Understanding contrastive representation learning through alignment
  and uniformity on the hypersphere.
\newblock In \emph{International Conference on Machine Learning}, pp.\
  9929--9939. PMLR, 2020.

\bibitem[Wu et~al.(2022)Wu, Gao, Lin, Han, Wang, and Hu]{wu2022infocse}
Wu, X., Gao, C., Lin, Z., Han, J., Wang, Z., and Hu, S.
\newblock Infocse: Information-aggregated contrastive learning of sentence
  embeddings.
\newblock \emph{arXiv preprint arXiv:2210.06432}, 2022.

\bibitem[Xie et~al.(2022)Xie, Sun, Liu, Wu, Gao, Zhang, Ding, and
  Cui]{xie2022contrastive}
Xie, X., Sun, F., Liu, Z., Wu, S., Gao, J., Zhang, J., Ding, B., and Cui, B.
\newblock Contrastive learning for sequential recommendation.
\newblock In \emph{2022 IEEE 38th international conference on data engineering
  (ICDE)}, pp.\  1259--1273. IEEE, 2022.

\bibitem[Yao et~al.(2020)Yao, Yi, Cheng, Yu, Menon, Hong, Chi, Tjoa, Kang, and
  Ettinger]{yao2020self}
Yao, T., Yi, X., Cheng, D.~Z., Yu, F., Menon, A.~K., Hong, L., Chi, E.~H.,
  Tjoa, S., Kang, J., and Ettinger, E.
\newblock Self-supervised learning for deep models in recommendations.
\newblock \emph{CoRR}, 2020.

\bibitem[Zhou et~al.(2020)Zhou, Wang, Zhao, Zhu, Wang, Zhang, Wang, and
  Wen]{zhou2020s3}
Zhou, K., Wang, H., Zhao, W.~X., Zhu, Y., Wang, S., Zhang, F., Wang, Z., and
  Wen, J.-R.
\newblock S3-rec: Self-supervised learning for sequential recommendation with
  mutual information maximization.
\newblock In \emph{Proceedings of the 29th ACM international conference on
  information \& knowledge management}, pp.\  1893--1902, 2020.

\end{thebibliography}
\bibliographystyle{icml2023}

\newpage
\appendix
\onecolumn

\section{$k_{learn}$ COLLAPSING PROBLEM AND REGULARIZATION}\label{App_sec:collapse_phenomenon}
In our setting, $k_{learn}$ is output of submodel $g$.
The objective of minimizing $\mathcal{L}_{learn}$ is to make the probability of positive pairs' similarity closer to 1.
However, a simple solution to achieve this is to make $M^{-}(i)$ an empty set, which means that ${\text{sim}(f(s_i), f(s_j))} \,>\, k_{learn}[i]$ for all pairs, resulting in every $k_{learn}$ value being set to -1 due to the cosine similarity.
Consequently, making the model $g$ output -1 becomes the easiest way to optimize $\mathcal{L}_{learn}$ loss.
To prevent this collapsing situation, we introduce a regularization term using $k_{stat}$.

\begin{equation}\label{eq:infoNCE_appendix}
\begin{alignedat}{3}
    \mathcal{L}_{learn} &=\sum \limits _{i=1}^{2B} \Bigg[-\log \cfrac{e^{\text{sim}(f(s_i), f(s_i^+))}}{\sum \limits _{j \in M^{-}(i)\cup\{i^+\}} e^{\text{sim}(f(s_i), f(s_j))}} \Bigg], \\
    M^{-}(i) &= \,\{\, j \in N(i)  \mid {\text{sim}(f(s_i), f(s_j))}\leq k_{learn}[i] \,\}, \\
    N(i) &= \{\, 1,2, ..., 2B-2 \, \}.
\end{alignedat}
\end{equation}

\section{Analysis For Similarity}
As shown in Fig \ref{fig:similarity}, both CL4SRec and DuoRec exhibit a concentration of similarities around the value 0.2. Since cosine similarity is used, we can infer that our feature points lie on an n-sphere denoted as $\textbf{S}^{n}$.
In $\textbf{S}^n$, the ideal scenario for points to be uniformly distributed is when most pairs of points are perpendicular to each other as below.
The maximum number of points that can have an angle equal or greater than $90^{\circ}$ between any two points is $2n+2$, indicating a situation where the points are most uniformly distributed on $\textbf{S}^n$.
In our experiments, we set $n$ to be 64, and the ML-1M dataset contains approximately 6000 points. Therefore, the value of 0.2, being greater than 0, appears to be an obvious result since there are numerous points compared to the dimension $n$. However, it may appear less uniform.
To address this, we experimented with higher values for $n$, such as 128, 256, and so on, aiming to achieve greater uniformity and improved metrics as a result.
We observed that as $n$ increased, the training reached saturation faster, but the final metric results decreased.

\begin{figure}[h]
    \centering
    \includegraphics[width=0.6\columnwidth]{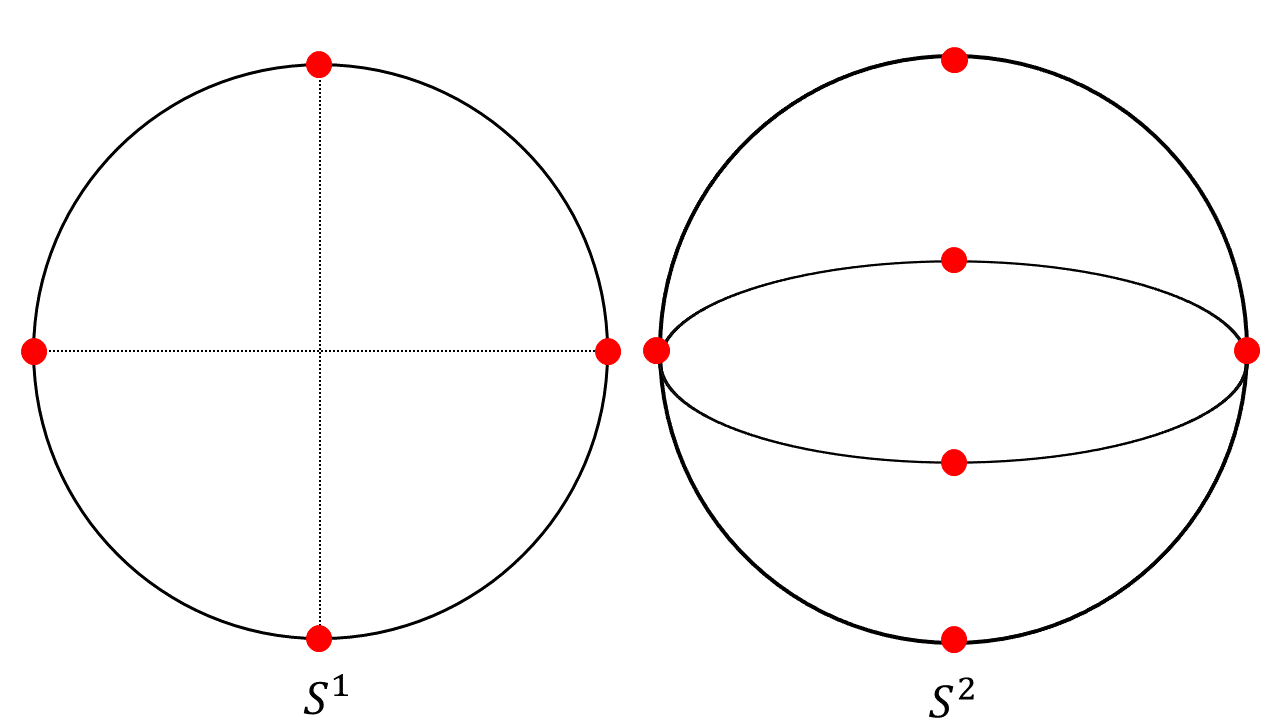}
    \label{fig:n-sphere}
\end{figure}

\section{Experimental Details}
We conduct all the experiments under the DuoRec\cite{qiu2022contrastive} github code.


\end{document}